\documentstyle[preprint,prl,aps,epsf]{revtex}

\newcommand{\bb}{$B^0$--$\bar{B}^0$}
\newcommand{\kk}{$K^0$--$\bar{K}^0$}
\newcommand{\ek}{$\varepsilon_K$}
\newcommand{\bsbs}{$B_s$--$\bar{B}_s$}
\newcommand{\bdbd}{$B_d$--$\bar{B}_d$}
\newcommand{\dmbs}{$\Delta m_s$}
\newcommand{\dmbd}{$\Delta m_d$}
\newcommand{\dmbsd}{$\Delta m_s/\Delta m_d$}
\newcommand{\bsg}{$b \to s\,\gamma$}
\newcommand{\meg}{$\mu \to e\,\gamma$}
\newcommand{\tmg}{$\tau \to \mu\,\gamma$}

\newcommand{\Bbsg}{$\text{B}(b \to s\,\gamma)$}
\newcommand{\Bmeg}{$\text{B}(\mu \to e\,\gamma)$}
\newcommand{\Btmg}{$\text{B}(\tau \to \mu\,\gamma)$}
\newcommand{\Bteg}{$\text{B}(\tau \to e\,\gamma)$}
\newcommand{\dlt}{$\delta_{13}$}

\begin{document}
\preprint{
\begin{tabular}{l}
KEK Preprint 99-176 \\
KEK-TH-677 \\
\rule{0em}{6ex}
\end{tabular}
}

\draft

\title{
Neutrino oscillation, SUSY GUT and $B$ decay
}

\author{
Seungwon Baek,$^1$
Toru Goto,$^1$
Yasuhiro Okada,$^{1,2}$
and
Ken-ichi Okumura$^{1,3}$
}

\address{
$^1$Theory Group, KEK, Tsukuba, Ibaraki, 305-0801 Japan
\\
$^2$Department of Particle and Nuclear Physics and
$^3$Department of Accelerator Science,\\
The Graduate University of Advanced Studies,
Tsukuba, Ibaraki, 305-0801 Japan
}

\date{February 14, 2000}

\maketitle

\begin{abstract}
Effects of supersymmetric particles on flavor changing neutral current
and lepton flavor violating processes are studied in the supersymmetric
SU(5) grand unified theory with right-handed neutrino supermultiplets.
Using input parameters motivated by neutrino oscillation, it is shown
that the time-dependent CP asymmetry of radiative $B$ decay can be as
large as 30\% when the \tmg\ branching ratio becomes close to the
present experimental upper bound.
We also show that the \bsbs\ mixing can be significantly different from
the presently allowed range in the standard model.
\end{abstract}

\pacs{}

Although the standard model (SM) of the elementary particle theory
describes current experimental results very well, particles and
interactions outside of the SM may appear beyond the energy scale
available at current collider experiments.
One of indications is already given by the atmospheric and the solar
neutrino anomalies which have been interpreted as evidences of
neutrino oscillation \cite{SKatm,SKsol}.
A natural way to introduce small neutrino masses for the neutrino
oscillation is the see-saw mechanism \cite{seesaw} where the
right-handed neutrino is introduced with a very heavy mass.
This scenario suggests the existence of a new source of flavor mixings
in the lepton sector at much higher energy scale than the electroweak
scale.

In this letter we consider flavor changing neutral current (FCNC)
processes and lepton flavor violation (LFV) of charged lepton decays in
the model of a SU(5) supersymmetric (SUSY) grand unified theory (GUT)
which incorporates the see-saw mechanism for the neutrino mass
generation.
In the SUSY theory, the superpartners of quarks and leptons, namely
squarks and sleptons respectively, have new flavor mixings in their mass
matrices.
In the model based on the minimal supergravity these mass matrices are
assumed to be flavor-blind at the Planck scale.
However renormalization effects due to Yukawa coupling constants can
induce flavor mixing in the squark/slepton mass matrices.
In the present model sources of the flavor mixing are Yukawa coupling
constant matrices for quarks and leptons as well as that for the
right-handed neutrinos.
Because the quark and lepton sectors are related by GUT interactions,
the flavor mixing relevant to the Cabibbo-Kobayashi-Maskawa (CKM) matrix
can generate the LFV such as \meg\ and \tmg\ processes \cite{LFV} in
addition to FCNC in hadronic observables \cite{FCNC}.
In the SUSY model with right-handed neutrinos, it is possible that the
branching ratios of the LFV processes become large enough to be measured 
in near-future experiments \cite{nuR-LFV}.
When we consider the right-handed neutrinos in the context of GUT, the
flavor mixing related to the neutrino oscillation can be a source of the
flavor mixing in the squark sector.
We show that due to the large mixing of the second and third generations
suggested by the atmospheric neutrino anomaly, \bsbs\ mixing, the
time-dependent CP asymmetry of the $B \to M_s\,\gamma$ process, where
$M_s$ is a CP eigenstate including the strange quark, can have a large
deviation from the SM prediction.

The Yukawa coupling part and the Majorana mass term of the
superpotential for the SU(5) SUSY GUT with right-handed neutrino
supermultiplets is given by
\begin{equation}
  W =
    \frac{1}{8}f_U^{ij} \Psi_{i} \Psi_{j} H_{5}
  + f_D^{ij} \Psi_{i} \Phi_{j} H_{\bar{5}}
  + f_N^{ij} N_{i} \Phi_{j} H_{5}
  + \frac{1}{2} M_{\nu}^{ij} N_{i} N_{j}
~,
\label{eq:SP}
\end{equation}
where $\Psi_{i}$, $\Phi_{i}$ and $N_{i}$ are ${\bf 10}$,
$\bar{\bf 5}$ and ${\bf 1}$ representations of SU(5) gauge group.
$i,j=1,2,3$ are the generation indices.
$H_{5}$ and $H_{\bar{5}}$ are Higgs superfields with ${\bf 5}$ and
$\bar{\bf 5}$ representations.
In terms of the SU(3)$\times$SU(2)$_L\times$U(1)$_Y$,
$\Psi_i$ contains $Q_i\,( {\bf 3},\, {\bf 2},\, \frac{1}{6} )$, 
$U_i\,( \bar{\bf 3},\, {\bf 1},\, -\frac{2}{3} )$ and
$E_i\,( {\bf 1},\, {\bf 1},\, 1 )$ superfields.
Here the representations for SU(3) and SU(2) groups and the U(1)$_Y$
charge are indicated in the parentheses.
$\Phi_i$ includes $D_i\,( \bar{\bf 3},\, {\bf 1},\, \frac{1}{3} )$ and
$L_i\,( {\bf 1},\, {\bf 2},\, -\frac{1}{2} )$, and $N_i$ is a singlet of 
SU(3)$\times$SU(2)$_L\times$U(1)$_Y$.
$M_\nu$ is the Majorana mass matrix.
Below the GUT scale ($\approx 2 \times 10^{16}$ GeV) and the Majorana
mass scale ($\equiv M_R$) the superpotential for the minimal
supersymmetric standard model (MSSM) fields is given by
\begin{equation}
  W_{\text{MSSM}} =
    \tilde{f}_U^{ij} Q_i U_j H_2
  + \tilde{f}_D^{ij} Q_i D_j H_1
  + \tilde{f}_L^{ij} E_i L_j H_1
  + \mu H_1 H_2
  - \frac{1}{2} \kappa_{\nu}^{ij} (L_i H_2)(L_j H_2)
~,
\label{Wmssm}
\end{equation}
where $\kappa_{\nu}$ is obtained by integrating out the heavy
right-handed neutrino fields.
At the right-handed neutrino mass scale $\kappa_{\nu}$ is given as
$\kappa_{\nu}^{ij} = ( f_N^{\bf T} M_{\nu}^{-1}f_N )^{ij}$.
The Yukawa coupling constants $\tilde{f}_U^{ij}$, $\tilde{f}_D^{ij}$
and $\tilde{f}_L^{ij}$ are related to the coupling constants $f_U^{ij}$
and $f_D^{ij}$ at the GUT scale.
The quark, charged lepton and neutrino masses and mixings are determined
from the superpotential Eq.\ (\ref{Wmssm}) at the low energy scale.

As discussed above, the renormalization effects due to the Yukawa
coupling constants induce various FCNC and LFV effects from the
mismatch between the quark/lepton and squark/slepton diagonalization
matrices.
In particular the large top Yukawa coupling constant is responsible for
the renormalization of the $\tilde{q}_L$ and $\tilde{u}_R$ mass
matrices.
At the same time the $\tilde{e}_R$ mass matrix receives sizable
corrections between the Planck and the GUT scales and various LFV
processes are induced.
In a similar way, if the neutrino Yukawa coupling constant $f_N^{ij}$ is
large enough, the $\tilde{l}_L$ mass matrix and the $\tilde{d}_R$ mass
matrix receive sizable flavor changing effects due to renormalization
between the Planck and the $M_R$ scales and the Planck and the GUT
scales, respectively.
These are sources of extra contributions to LFV processes and various
FCNC processes such as \bsg, the \bb\ mixing and the \kk\ mixing.

It is particularly interesting that the chiral structure of the FCNC
amplitudes due to the $\tilde{d}_R$ mixing is different from that
expected in the SM.
For example, the flavor mixing in the $\tilde{d}_R$ sector generates a
sizable contribution to the $b \to s\,\gamma_R$ amplitude through
gluino--$\tilde{d}_R$ loop diagrams, whereas
this amplitude is suppressed by a factor $m_s/m_b$ over the dominant
$b \to s\,\gamma_L$ amplitude in the SM.
When the amplitudes with both chiralities exist, the mixing-induced
time-dependent CP asymmetry in the $B \to M_s\,\gamma$ process can be
induced.
Using the Wilson coefficients $c_7$ and $c'_7$ in the effective
Lagrangian for the \bsg\ decay
$
{\cal L} = c_7 \bar{s} \sigma^{\mu\nu} b_R F_{\mu\nu}
          +c'_7 \bar{s} \sigma^{\mu\nu} b_L F_{\mu\nu} + \text{H.c.}
$,
the asymmetry is written as
\begin{displaymath}
  \frac{\Gamma(t) - \bar{\Gamma}(t)}{\Gamma(t) + \bar{\Gamma}(t)} =
  \xi A_t \sin\Delta m_d t
~,
~~~
   A_t =
   \frac{2\text{Im}({\rm e}^{-i\theta_B} c_7 c'_7)}{|c_7|^2 + |c'_7|^2}
~,
\end{displaymath}
where $\Gamma(t)$ ($\bar{\Gamma}(t)$) is the decay width of
$B^0(t) \to M_s\,\gamma$ ($\bar{B}^0(t) \to M_s\,\gamma$) and $M_s$ is some
CP eigenstate ($\xi = +1(-1)$ for a CP even (odd) state) such as
$K_S\, \pi^0$ \cite{AGS-CHH}.
$\Delta m_d = 2|M_{12}(B_d)|$ and $\theta_B = \arg M_{12}(B_d)$ where
$M_{12}(B_d)$ is the \bdbd\ mixing amplitude.
Because the asymmetry can be only a few percent in the SM, a sizable
asymmetry is a clear signal of new physics beyond the SM.

We calculated the following observables in the FCNC and LFV processes:
the CP violation parameter in the \kk\ mixing \ek, \bdbd\ and \bsbs\
mass splittings \dmbd\ and \dmbs, respectively, $A_t$
and the branching ratios \Bbsg, \Bmeg, \Btmg\ and \Bteg.
We solved renormalization group equations (RGEs) for Yukawa coupling
constants and the SUSY breaking parameters numerically keeping all
flavor matrices.
After demanding the condition of radiative electroweak symmetry
breaking, the free parameters in the minimal supergravity model are the
universal scalar mass $m_0$, the universal gaugino mass $M_0$, the
scalar trilinear parameter $A_0$, the ratio of two vacuum expectation
values $\tan\beta$ and the sign of the Higgsino mass parameter $\mu$.
In addition we need to specify neutrino parameters.
The phenomenological inputs from neutrino oscillation are two
mass-squared differences and the Maki-Nakagawa-Sakata (MNS) matrix.
In order to relate these parameters to $f_N$ and $M_\nu$, we work in
the basis for $N_i$, $L_i$ and $E_i$ where
$\tilde{f}_L^{ij}=\hat{f}_L^{ij}$ and $f_N^{ij}=(\hat{f}_N V_L)^{ij}$
($\hat{f}_L$ and $\hat{f}_N$ are diagonal matrix) at the matching scale
$M_R$.
In this basis
$
\kappa_\nu = V_L^{\bf T}\hat{f}_N M_\nu^{-1} \hat{f}_N V_L
= V_{\text{MNS}}^{0*} \hat{\kappa}_\nu V_{\text{MNS}}^{0\dagger}
$
where $V_{\text{MNS}}^0$ is the MNS matrix at $M_R$ and
$\hat{\kappa}_\nu$ is a diagonal matrix.
Note that although $V_L=V_{\text{MNS}}^{0\dagger}$ when $M_\nu$ is
diagonal in this basis, two are independent in a general case.
Once we specify three neutrino masses, $V_{\text{MNS}}$, $V_L$ and
$\hat{f}_N$ we can determine the $M_\nu$ matrix.
Then using the GUT relation for Yukawa coupling constants, we can
calculate all squark and slepton mass matrices through RGEs.
Note that $V_L$ essentially determines the flavor mixing in the
$\tilde{d}_R$ and $\tilde{l}_L$ sectors in this basis.

As typical examples of the neutrino parameters, we consider the
following parameter sets corresponding to (i) the
Mikheyev-Smirnov-Wolfenstein (MSW) small mixing
angle and (ii) the MSW large mixing angle solutions for the solar
neutrino problem \cite{SKsol}.
\\
(i) small mixing:
\begin{eqnarray}
  m_{\nu} &=&
  \left(
    2.236 \times 10^{-3},\,
    3.16 \times 10^{-3},\,
    5.92 \times 10^{-2}
  \right) ~{\rm eV}
~,
\nonumber\\
  V_{\text{MNS}} &=&
  \left(
    \begin{array}{ccc}
         0.999
      &  0.0371
      &  0
      \\ -0.0262
      &  0.707
      &  0.707
      \\ 0.0262
      &  -0.707
      &  0.707
    \end{array}
  \right)
~,
\nonumber
\end{eqnarray}
(ii) large mixing:
\begin{eqnarray}
  m_{\nu} &=&
  \left(
    4.0 \times 10^{-3},\,
    5.831 \times 10^{-3},\,
    5.945 \times 10^{-2}
  \right)~{\rm eV}
~,
\nonumber\\
  V_{\text{MNS}} &=&
  \left(
    \begin{array}{ccc}
         1/\sqrt{2}
      &  1/\sqrt{2}
      &  0
      \\ -1/2
      &  1/2
      &  1/\sqrt{2}
      \\ 1/2
      &  -1/2
      &  1/\sqrt{2}
    \end{array}
  \right)
~.
\nonumber
\end{eqnarray}
In each example we also take $M_{\nu}$ to be proportional to a unit
matrix with a diagonal element of $M_R = 4 \times 10^{14}$ GeV so that
$V_L = V_{\text{MNS}}^{0\dagger}$ and
$\hat{f}_N^{ii} = \sqrt{M_R \hat{\kappa}_\nu^{ii}}$.
We fix $m_t^{\text{pole}}=175$ GeV, $m_b^{\text{pole}}=4.8$ GeV and the
CKM parameters as $V_{cb}=0.04$, $|V_{ub}/V_{cb}|=0.08$ and take several
values of the phase parameter in the CKM matrix \dlt\ \cite{CK-PDG}.
We take $\tan\beta=5$ and vary other SUSY parameters $m_0$, $M_0$, $A_0$
and the sign of $\mu$.
Various phenomenological constraints from SUSY particles search are
included (for detail see \cite{Goto}).
We also impose
$2 \times 10^{-4} < \text{B}(b\to s\,\gamma) < 4.5 \times 10^{-4}$
\cite{bsgExp} in the following analysis.

\begin{figure}
\begin{center}
\rule{0em}{24ex}\\
\makebox[0em]{
\def\epsfsize#1#2{0.75#1}
\epsfbox{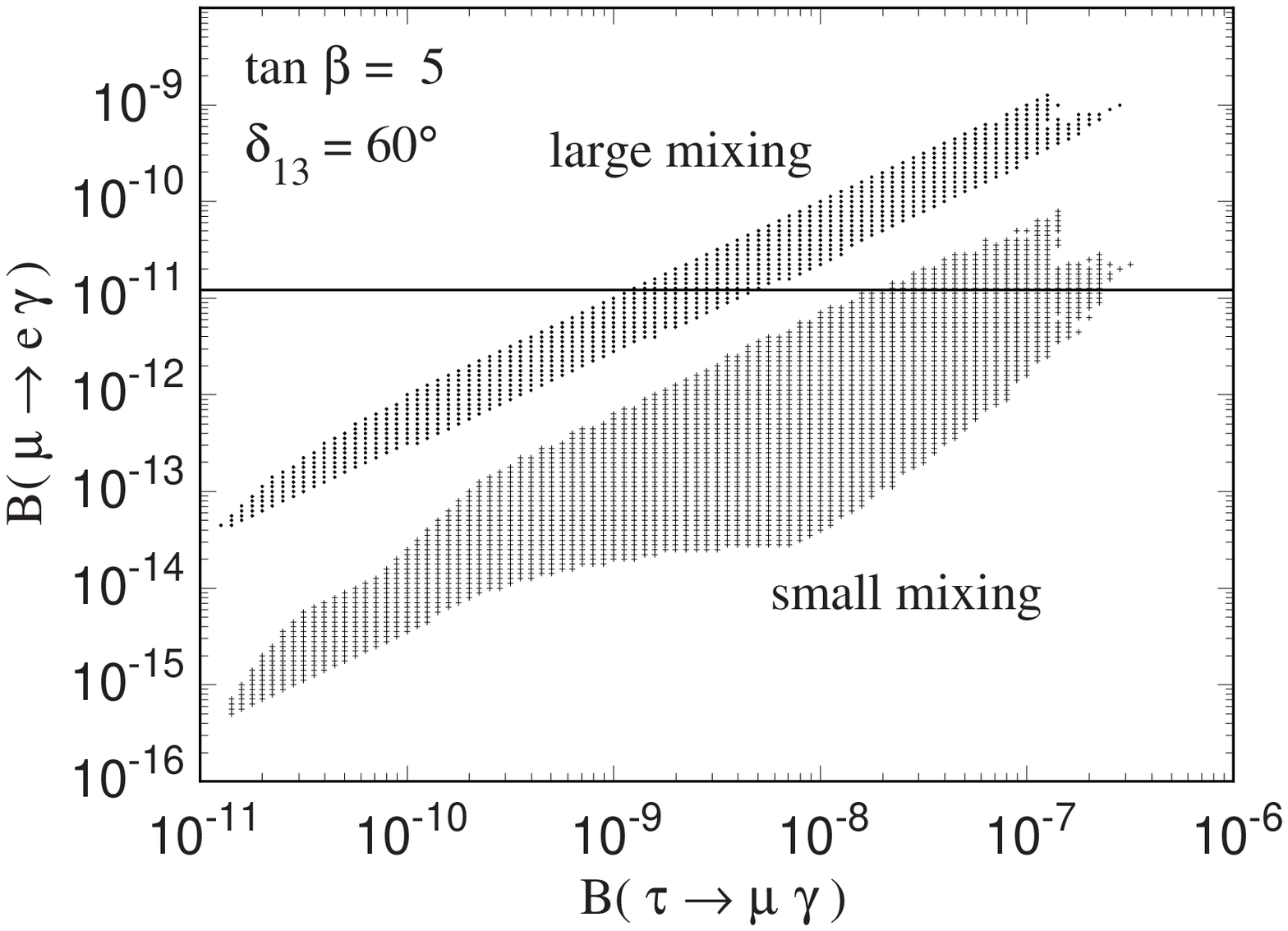}
}
\end{center}
\caption{
The correlation between \Bmeg\ and \Btmg\ for small and large mixing
angle cases.
See text for the input parameters.
}
\label{fig:meg-tmg}
\end{figure}

Let us first discuss \Bmeg\ and \ek\ which turn out to be strong
constraints on the parameter space in this model.
Fig.~\ref{fig:meg-tmg} shows the correlation between \Bmeg\ and \Btmg\
for the neutrino parameter set (i) and (ii) for
$\delta_{13} = 60^\circ$.
We can see that the \Bmeg\ becomes a very strong constraint for the case
(ii), which is a reflection of the large 1--2 mixing in the
$V_{\text{MNS}}$ matrix.
By requiring
$\text{B}(\mu \to e\,\gamma) < 1.2 \times 10^{-11}$
\cite{megExp}, \Btmg\ becomes less than $10^{-8}$ for the case (ii)
whereas it can be close to the present experimental bound
($1.1\times 10^{-6}$ \cite{tmgExp}) for the case (i).
We also calculated \Bteg\ which turns out to be smaller than
$3\times 10^{-12}$ in both cases.
The constraint from \ek\ depends on the parameter \dlt.
After imposing the \Bmeg\ constraint, \ek\ can be enhanced by 50\% for
the case (i) and by a factor of 2 for the case (ii).
This means that compared to favorable values in the SM
($50^\circ < \delta_{13} < 90^\circ$), a smaller value of \dlt\ 
is allowed due to the extra contributions.

\begin{figure}
\begin{center}
\makebox[0em]{
\def\epsfsize#1#2{0.75#1}
\epsfbox{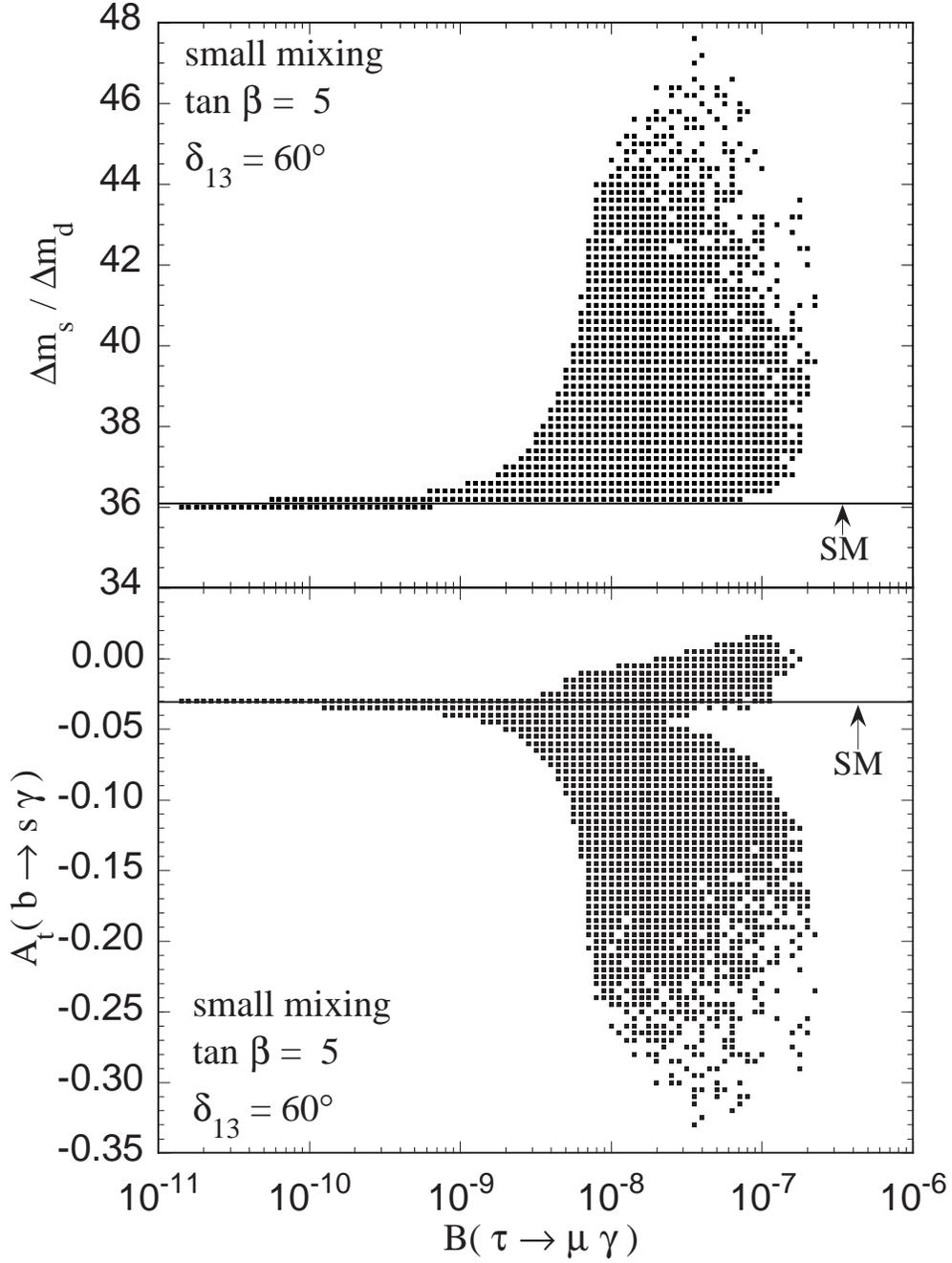}
}
\end{center}
\caption{
The ratio of \bsbs\ and \bdbd\ mass splittings \dmbsd\ and
the magnitude factor $A_t$ of the time-dependent CP asymmetry in the
$B\to M_s\,\gamma$ process as a function of \Btmg\ for the small mixing
case (i).
}
\label{fig:dmbsdAt-tmg}
\end{figure}

The upper part of Fig.~\ref{fig:dmbsdAt-tmg} shows a correlation between
\dmbsd\ and \Btmg\ for case (i) and $\delta_{13}=60^{\circ}$.
Here we imposed the constraints from \Bmeg\ and \ek.
We also imposed the constraint from \dmbd\ itself though the deviation
of this quantity from the SM value is within 5\%.
For the theoretical uncertainties we allow $\pm15$\% difference for \ek\
and $\pm40$\% for \dmbd.
For \dmbsd\ we fix the hadronic parameters as $f_{B_s}/f_{B_d}=1.17$ and 
$B_{B_s}/B_{B_d}=1$.
We can see that \dmbsd\ can be enhanced up to 30\% compared to the SM
prediction.
This feature is quite different from the minimal supergravity model
without the GUT and right-handed neutrino interactions \cite{Goto} where
\dmbsd\ is almost the same as the SM value.
$A_t$ for the same parameter
set is shown as a function of \Btmg\ in the lower part of
Fig.~\ref{fig:dmbsdAt-tmg}.
We can see that $|A_t|$ can be close to 30\% when \Btmg\ is larger than
$10^{-8}$.
The large asymmetry arises because the renormalization effect due to 
$f_N$ induces sizable contribution to $c'_7$ through
gluino--$\tilde{d}_R$ loop diagrams.
The corresponding figure to Fig.~\ref{fig:dmbsdAt-tmg} for the case
(ii) shows that \Btmg\ is cut off below $10^{-8}$ and the maximal
deviation of \dmbsd\ from SM is within 6\% and $|A_t|$ becomes at most 6\%.

\begin{figure}
\begin{center}
\makebox[0em]{
\def\epsfsize#1#2{0.75#1}
\epsfbox{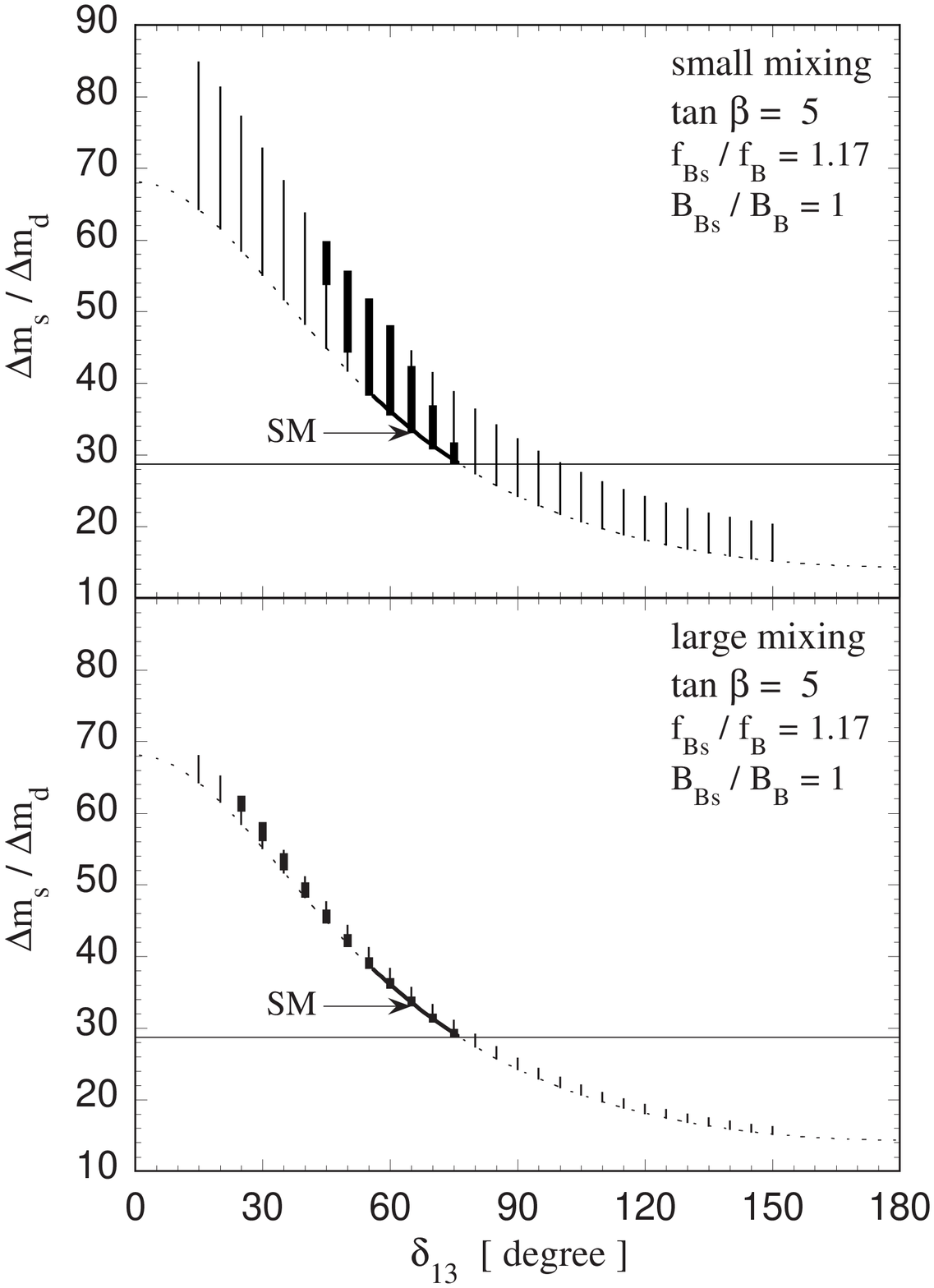}
}
\end{center}
\caption{
Possible range of \dmbsd\ as a function of \dlt\ for small and large
mixing cases.
Each thick vertical lines shows allowed range with the experimental
constraints (see text).
Thin vertical lines correspond to the case without the constraints from
\ek, \dmbd\ and \dmbsd.
Horizontal line shows the experimental lower bound of \dmbsd.
The dotted line corresponds to the SM value and the solid section shows
the allowed range in the SM.
}
\label{fig:dmbsd-dlt}
\end{figure}

In Fig.~\ref{fig:dmbsd-dlt} we show \dmbsd\ for several values of \dlt\
for the case (i) and (ii).
In these figures we impose \Bmeg\ constraint.
Thin vertical lines
correspond to the case without the experimental constraints from \ek,
\dmbd\ and the lower bound for \dmbsd\ \cite{dmbsExp},
and the thick lines are allowed range with these constraints.
The allowed range of the SM is also shown in these figures.
Because the new contributions to \bdbd\ amplitude is small, the
time-dependent CP asymmetry of $B\to J/\psi\,K_S$ in this model is
essentially the same as the SM value, therefore we can obtain
information on \dlt\ once this asymmetry is measured
experimentally.
For example the asymmetry of $B\to J/\psi\,K_S$ mode is 0.4 for
$\delta_{13}=25^\circ$ and 0.65 for $\delta_{13}=75^\circ$.
This figure means the possible deviation from the SM may be seen
in both cases once the CP asymmetry of $B\to J/\psi\,K_S$ mode and
\dmbsd\ are measured.

In our example we took $M_R$ which corresponds to the upper bound of
$f_N$ because a larger $M_R$ would lead to the blow-up of $f_N$ below
the Planck scale.
If we take a lower value of $M_R$ the flavor changing amplitudes
essentially scale as $M_R$.

Finally we would like to comment on genarizations of our result.
Firstly for a large $\tan\beta$ case, the constraint from \Bmeg\ becomes
stronger.
As a result we cannot see any deviation from the SM for the large
mixing case in the figures corresponding to Fig.\
\ref{fig:dmbsdAt-tmg} and \ref{fig:dmbsd-dlt} for $\tan\beta=30$,
whereas the result is similar for the small mixing case.
Secondly we considered the vacuum oscillation case.
We see that the pattern of the deviation from the SM is similar to the
small mixing case because the effect of the flavor mixing in 1--2
generations turns out to be suppressed by the degeneracy of the two
light neutrino masses.
In this letter we took the case where $M_\nu$ is proportional to a unit
matrix.
When we consider a more general case, $V_L$ is not necessarily equal to
$V_{\text{MNS}}^{0\dagger}$.
In addition, the superpotential (\ref{eq:SP}) does not lead to the
realistic fermion mass relation especially for the first and the second 
generations.
In order to solve this problem we may have to introduce more free
parameters.
Although the precise values of the predictions depend on the detail of
the model in question, the possible new physics signals such as \Btmg,
\bsbs\ mixing and $A_t$ may be expected as long as some of the neutrino
Yukawa coupling constants are large.
Because these signals provide quite different signatures compared to
the SM and the minimal supergravity model without GUT and right-handed
neutrino interactions, future experiments in $B$ physics and LFV can
provide us important clues on the interactions at very high energy scale.

S.\ B.\ would like to thank KOSEF for financial support.
The work of Y.\ O.\ was supported in part by the Grant-in-Aid of the
Ministry of Education, Science, Sports and Culture, Government of Japan
(No.\ 09640381), Priority area ``Supersymmetry and Unified Theory of
Elementary Particles'' (No.\ 707), and ``Physics of CP Violation''
(No.\ 09246105).

\end{document}